\documentstyle[epsf,12pt]{article}

\begin{document}

\begin{titlepage}
\begin{flushright}
%\preprint{
OSU-HEP-02-05\\
UH-511-998-02\\
April 2002\\
%}
\end{flushright}
\vskip 2cm
\begin{center}
%\title
{\Large\bf
Lepton Number Violating Muon Decay \\}
\vspace*{0.1in}
{\Large\bf and the LSND Neutrino Anomaly\\}
\vskip 1cm
{\large\bf K.S.\ Babu$\,{}^1$ and Sandip Pakvasa$\,{}^{2}$} \\
\vskip 0.5cm
%\address
{\it ${}^1\,$Department of Physics, Oklahoma State University\\
Stillwater, OK~~74078, USA\\ [0.1truecm]
${}^2\,$Department of Physics and Astronomy, University of Hawaii\\
Honolulu, HI~~96822, USA\\[0.1truecm]
}

%\maketitle
\end{center}
\vskip .5cm

\begin{abstract}

We show that lepton number violating muon decays, $\mu^+ \rightarrow
e^+ + \overline{\nu_e} + \overline{\nu_i}$ ($i=e, \mu$ or $\tau$), can consistently
explain the neutrino anomaly reported by the LSND experiment.  Two
effective operators in the Standard Model are identified which lead to just such decays
and no other processes. The scale of new physics $\Lambda$  must be relatively low,
$\Lambda \leq 500$ GeV. Extensions of the Standard Model which realize these
effective operators are presented. Since new physics affects only the decay
of the muon, and not of $\pi^\pm$, our scenario predicts a null result for
$\overline{\nu}_\mu-\overline{\nu}_e$
oscillation searches at the Fermilab mini--BOONE  experiment.
Models which realize these effective operators, while consistent with all
available data, can be tested in the near future through (i) discovery of
new scalar particles with masses below about 500 GeV,
(ii) small but observable deviations in $e^+  e^- \rightarrow \mu^+ \mu^-$ and
$e^+ e^- \rightarrow \nu \nu \gamma$ cross sections, (iii) observable corrections
to the muon $g-2$, and (iv) lepton number violating $Z^0$ decays with branching
ratios of order $10^{-7}$.

\end{abstract}

\end{titlepage}

\section{Introduction}

Evidence for neutrino flavor oscillations have been mounting over the years.
A variety of solar neutrino experiments \cite{cl,superk,gallex,sage,sno}
which have consistently detected  fewer  $\nu_e$'s from the sun than
expected now seem to converge on the large angle MSW solution
as the preferred explanation for the discrepancy \cite{bahcall}.  The deficit in
the flux of atmospheric muon neutrinos \cite{atm} and especially the observed zenith
angle dependence are compelling evidences in favor of $\nu_\mu-\nu_\tau$
oscillations.  The LSND collaboration observes an anomaly in the flux of
$\overline{\nu}_e$ detected \cite{lsnd}, which can be interpreted as evidence for
$\overline{\nu}_\mu-\overline{\nu}_e$ oscillations.

As it turns out, it is not possible to explain all of these observations
in terms of neutrino oscillations with just three flavors of neutrinos
($\nu_e,~\nu_\mu$ and $\nu_\tau$).  The reason is that the
characteristic oscillation length scales inferred from the three types
of experiments are quite different having no overlap: $\lambda_\odot \sim
200~{\rm km},~\lambda_{atm} \sim 6000 ~{\rm km}$ and $\lambda_{LSND} \sim
60$ m. The oscillation length is given by $\lambda = (4E)/\Delta m^2$ where
$\Delta m^2_{12} = m_1^2-m_2^2$ etc, with $m_i$ being the mass of the $i$th
neutrino mass eigenstate, and where $E$ denotes the neutrino energy.
In three neutrino oscillation schemes there are only two independent mass--splittings,
which therefore excludes a simultaneous explanation of all three observations.

Two approaches have been adopted in the literature to address this conundrum.
In the first, one of the three observations is simply discarded.  In the second
approach, a fourth ``sterile'' neutrino ($\nu_s$) -- sterile so that it does not affect
the number of neutrinos that couples to the $Z^0$ bososn -- is introduced.
Both of these approaches are not very satisfactory.  The first one is clearly without
justification.  The second, introduction of a sterile neutrino, has some serious theoretical
difficulties, in addition to a possible  problem with the standard
Big Bang Nucleosynthesis.
Its lightness cannot be explained in any simple way, unlike that of
the ``active'' neutrino which follows very naturally from the
seesaw mechanism \cite{seesaw}.  Even taking a light sterile neutrino for
granted, some potential experimental difficulties have emerged over the last
year.  When the recent SNO results on solar neutrinos are combined with
the results of SuperKamiokande, the allowed parameter space is tightly constrained
for oscillations involving a sterile neutrino.
In fact, a two--flavor $\nu_e-\nu_s$ oscillation scenario no longer
provides a good fit to the solar neutrino data.  Atmospheric
neutrino oscillation results from SuperKamiokande also disfavor oscillations
of $\nu_\mu$ with a $\nu_s$.  Somewhat more involved four neutrino oscillation
schemes would be necessary, conforming to either a (3+1) \cite{3p1} scheme or a
mixed 2+2 scheme \cite{mixed}.  Although both appear to be viable presently,
neither gives a really satisfactory global fit to all the data \cite{valle}.
There is also a proposal to account for the LSND results in the three neutrino
framework by invoking CPT violation \cite{murayama}.

The purpose of this paper is to provide a simultaneous solution to the three
neutrino anomalies without introducing a sterile neutrino.  It will be achieved
by small non--standard interactions of the leptons instead.  Since the
flavor conversion probabilities are of order unity for both the solar and
the atmospheric neutrino experiments, we shall focus on non--standard lepton
interactions explaining the LSND anomaly, which calls for a smaller probability:
$P_{\overline{\nu}_\mu\rightarrow \overline{\nu}_e} =
(0.264 \pm 0.067 \pm 0.045)\%$ \cite{lsnd}.
Specifically, we shall show that lepton number violating $(\Delta L = 2)$
muon decays can account for the LSND events.  The new decay mode of the
muon is
\begin{equation}
\mu^+ \rightarrow e^+ + \overline{\nu}_e + \overline{\nu}_i~,
\end{equation}
where $\overline{\nu}_i$ stands for any one of $\overline{\nu}_e,~\overline{\nu}_\mu$
or $\overline{\nu}_\tau$.  Consistent gauge models
will be presented where such decays occur with the desired strength.

It is crucial that the non--standard decays of the muon are lepton number
violating \cite{pak}.  Attempts to explain the LSND data in terms of lepton number
conserving four--fermion interaction \cite{grossman,grossman1} will lead to inconsistencies with
other experiments.  One can make an $SU(2)_L$ transformation on such effective
operators which would invariably generate lepton number violating processes
such as $\mu \rightarrow 3e$ decay, muonium--antimuonium transition,
$\tau\rightarrow \mu e e$ decay,
etc, which are highly constrained by experiments \cite{pak,grossman}.
In contrast, the $\Delta L = 2$
processes that we find will not admit an $SU(2)_L$ transformation to generate
one of these highly forbidden processes, primarily because there is no charged
current counterpart to the anomalous muon decays.\footnote{We shall see shortly
that the anomalous muon decay is mediated by neutral current.}
Since the new physics affects only $\mu^\pm$ decays, and not that of $\pi^\pm$,
our interpretation of the LSND results would predict that no oscillation signal
should be seen at the Fermilab mini--BOONE experiment.
We have identified two $\Delta L = 2$
effective operators which are invariant under the Standard Model
symmetries that can induce the decay of Eq. (1) without generating any other effects.
These effective operators can be realized from underlying renormalizable theories
which are extensions of the Standard Model involving additional scalar multiplets.
The masses of these scalar bosons are bounded by about 500 GeV, in order that the signal at
LSND is significant.  Since the scale of new physics, $\Lambda$, is relatively low,
the particles at the scale $\Lambda$ do not decouple entirely
and have an impact on low energy observables.
We highlight the processes that would succumb to the new physics most easily.
They include small but observable deviations in $e^+  e^- \rightarrow \mu^+ \mu^-$ and
$e^+ e^- \rightarrow \nu \nu \gamma$ cross sections, observable corrections
to the muon $g-2$, and lepton number violating $Z^0$ decays with branching
ratios of order $10^{-7}$.

\section{Lepton number violating muon decay}

The decay $\mu^+ \rightarrow e^+ + \overline{\nu}_e + \overline{\nu}_i$ with
$i=e~\mu$ or $\tau$, if allowed by the theory, can explain the LSND neutrino
anomaly.  Recall that the usual $\mu$ decay $\mu^+ \rightarrow e^+ + \overline{\nu}_\mu+
\nu_e$ has no $\overline{\nu}_e$ which the LSND detector registers.  The new decay
of $\mu^+$ produces $\overline{\nu}_e$, which will then be detected.  The branching
ratio for the new decay should be about $(1.5-3) \times 10^{-3}$, as can be inferred
from the LSND analysis of the $\overline{\nu}_\mu-\overline{\nu}_e$ oscillation
probability.

We have found two effective operators invariant under the symmetries of the
Standard Model \cite{bl} that can induce the desired $\Delta L = 2$ muon decays.  They are
\begin{eqnarray}
{\cal O}_1 &=& {1 \over \Lambda^5} (\overline{\Psi}_{\mu} e_R \Phi)(\Psi_e^T C^{-1} \Psi_i
\Phi \Phi) \nonumber \\
{\cal O}_2 &=& {1 \over \Lambda^5} (\overline{\mu}_R \Psi_e \Phi^\dagger) (\Psi_e^T C^{-1} \Psi_i
\Phi \Phi)~.
\end{eqnarray}
Here $\Psi_\mu = (\nu_\mu,~\mu)_L^T$, $\Psi_e=(\nu_e,~e)_L^T$, etc denote the left--handed
lepton doublets,
while $e_R,\mu_R$ denote the right--handed singlets.  $\Phi$ is the Standard Model
Higgs doublet with its hypercharge normalized to be $+1/2$.

Note that both ${\cal O}_{1}$ and ${\cal O}_{2}$ are non--renormalizable
operators of dimension 9, and hence are suppressed by fifth power of
$\Lambda$ which characterizes the scale of new physics.
As we shall see, these
operators will arise from integrating out scalar fields which have lepton number
violating interactions.  Operator ${\cal O}_1$ has a unique $SU(2)_L$ contraction,
owing to Bose symmetry in the $\Phi$ field.
Although $(\overline{\Psi}_\mu e_R \Phi)$ in ${\cal O}_1$ contracts to form an
$SU(2)_L$ singlet, it will not induce a $\overline{\mu}_L e_R$ mass term (see below).
Similarly, no $d=5$ neutrino mass term will arise from these operators.   Note also
that there is a unique Lorentz contraction of fermionic fields in ${\cal O}_1$,
if we limit ourselves to obtaining these operators by integrating out
scalar fields.  Similar remarks apply to operator ${\cal O}_2$ as well.  In this
case, there is a minor variant, obtained by contracting $(\overline{\mu}_R \Psi_i
\Phi^\dagger)$ rather than $(\overline{\mu}_R \Psi_e \Phi^\dagger)$, which is
somewhat different for the case when $i\neq e$.

When the vacuum expectation value $\left\langle \Phi^0\right \rangle = v$
is inserted in Eq. (2),
it would generate the four--fermion operators
\begin{eqnarray}
{\cal O}_1 &\sim& {v^3 \over \Lambda^5}(\overline{\mu}_L e_R)(\nu_e^T C^{-1} \nu_i) \nonumber \\
{\cal O}_2 &\sim& {v^3 \over \Lambda^5}(\overline{\mu}_R e_L)(\nu_e^T C^{-1} \nu_i)~.
\end{eqnarray}
These operators will lead to the decay $\mu^+ \rightarrow e^+ + \overline{\nu}_e +
\overline{\nu}_i$.
The branching ratio for the $L$--violating $\mu^+$ decay is given by
\begin{equation}
Br(\mu^+ \rightarrow e^+ \overline{\nu}_e \overline{\nu}_i) =
\left[{v^3 \over 4\sqrt{2}G_F \Lambda^5}\right]^2~.
\end{equation}
Inserting the numerical value of $v\sim 246$ GeV, and demanding that the branching
ratio for this decay is in the range $Br(\mu^+ \rightarrow e^+ \overline{\nu}_e
\overline{\nu}_i) = (0.0015-0.0025)$,
we find that the scale $\Lambda \simeq (360-340)$ GeV.
Since $\Lambda$ is not very large, the particles that have masses of order $\Lambda$
will not entirely decouple and will affect low energy observables.

An important point about operators ${\cal O}_1$ and ${\cal O}_2$ of
Eq. (2) is that they induce only the terms given in Eq. (3) and nothing more.
This is because of the $\Delta L =2$ nature of these operators and the fact that
there is only one  Higgs scalar in the theory.  This proves that the effective
operators of Eq. (2) generate the desired interactions that can explain the LSND
events, and nothing else.  Unlike in the case of $\Delta L = 0$ muon decays,
there is no conflict with processes such as $\mu \rightarrow 3 e$, $\tau \rightarrow
\mu e e$, etc.

KARMEN experiment \cite{karmen} has set severe limits on lepton number violating decays of the
muon.  These limits however, depend sensitively on the assumed decay distribution.
The anomalous muon decay arising from Eq. (2) in our models has a different distribution
compared to that of the usual (V-A) decay of the muon.  We find that operators
${\cal O}_1$ and ${\cal O}_2$ lead to the prediction $\rho = 0$ for the Michel
parameter for the $\Delta L = 2$ decay.  (The other decay asymmetry parameters
for the $L$--violating $\mu$ decay are found to be: $\eta=0,~ \xi = -3/4,$ and
$\delta = 0$ \cite{kuno}.)
The limit from KARMEN for the branching ratio
for the decay $\mu^+ \rightarrow e+\overline{\nu}_e+\overline{\nu}_i$
corresponding to the case where $\rho = 0$ is Br $\leq (0.0015-0.002)$ \cite{eitelcom}.
This limit is somewhat weaker than KARMEN's published limit corresponding to the
case of $\rho = 3/4$ \cite{eitel}.  Thus, the KARMEN results appear to be just
about consistent with LSND observations, in our interpretation of the data.
A joint analysis of the LSND and the KARMEN data in our framework will be
desirable.

All new phenomena in the model will arise from
particles with masses of order $\Lambda$.  To see these additional new effects,
we have to generate operators of Eq. (2) from renormalizable Lagrangian densities.
We now turn to this task.

\section {Gauge models for Lepton number violating muon decays}

We shall now present renormalizable gauge models that induce the effective operators
of Eq. (2).  These models are obtained by extending the scalar sector of the Standard
Model.  The new interaction Lagrangian for the first model which induces ${\cal O}_1$ is given by
\begin{eqnarray}
{\cal L}_1 &=& h_{\mu e}\overline{\Psi}_\mu e_R H + f_{ei} \Psi_e^T C^{-1} \Psi_i\Delta
\nonumber \\
&+& {\cal M}_0 H^\dagger \chi \Phi + \lambda'  \Delta \chi \Phi^\dagger \Phi^\dagger + H.C.
\end{eqnarray}
Here $\Delta(3,1),~\chi(3,0)$ and $H(1,1/2)$ are scalar fields (their $SU(2)_L \times
U(1)_Y$ quantum numbers are as indicated) which do not acquire any vacuum expectation value.

Figure 1 will induce the effective operator ${\cal O}_1$ from Eq. (5).  Notice that
the intermediate scalar particles are electrically neutral, so the anomalous muon decay
is a neutral current process.  It is not possible to make an $SU(2)_L$ transformation
in Fig. 1 to convert some of the external fermions, without also transforming the external
Higgs field to its charged component, which of course is unphysical.

%%%%%%%%%%%%%%%%%%%%%%%%%%%%%%%%%%%%%%%%%%%%%%%%%%%%%%%%%%%%%%%%%%%
\begin{figure}[h]
\centering
\epsfysize=2.0in
\hspace*{0in}
\epsffile{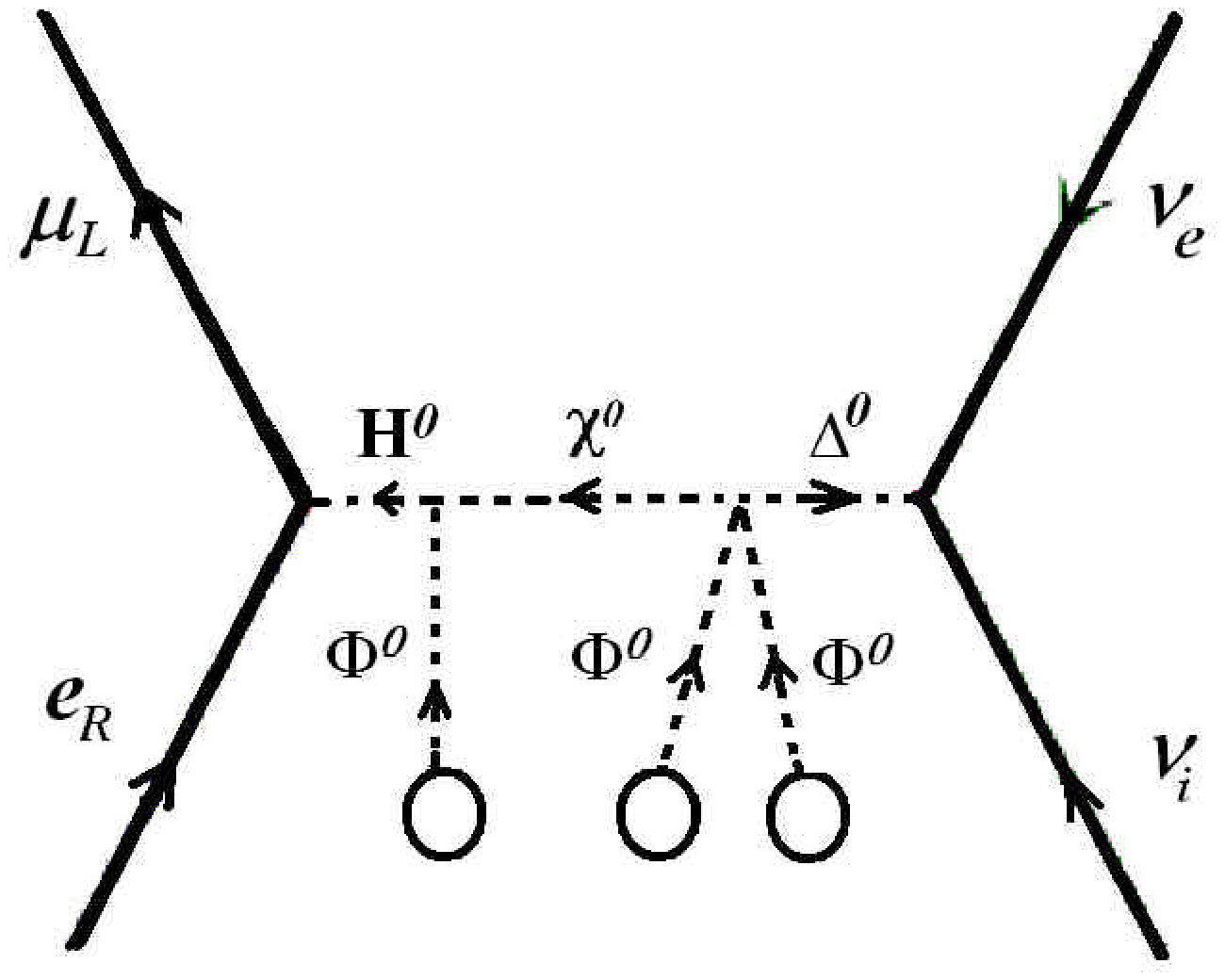}
\caption{Scalar exchange inducing lepton number violating decay
$\mu^+ \rightarrow e^+ + \overline{\nu}_e + \overline{\nu}_i$ through operator ${\cal O}_1$.}
\label{fig1:eps}
\end{figure}
%%%%%%%%%%%%%%%%%%%%%%%%%%%%%%%%%%%%%%%%%%%%%%%%%%%%%%%%%%%%%%%%%%%

The strength of the effective four--fermion coupling arising from Fig. 1, in
the approximation of small scalar mixing, is
\begin{equation}
G_{\rm eff} = {h_{\mu e} f_{ei} \lambda' {\cal M}_0 v^3 \over (m_{\chi^0}^2 m_{H^0}^2
m_{\Delta^0}^2)}~.
\end{equation}
If all the mass parameters are equal and if $h_{\mu e}, f_{ei}$ and $\lambda'$ are
equal to one, LSND data would require the scalar masses to be
in the range $(340-360)$ GeV.  Actually, we can make this statement more precise.
Let us denote the scalar mass--squared matrix involving the mixing of
$(H^0, \chi^0, \Delta^{0*})$ to be ${\cal M}^2$.  Let $K$ be the unitary matrix
that diagonalizes ${\cal M}^2$:  $K^\dagger {\cal M}^2 K = diag.(M_1^2,~M_2^2,~
M_3^2)$ with $M_1 \leq M_2 \leq M_3$.
Then by making use of the unitarity of $K$ and the fact that
$({\cal M}^2)_{13}=0$, we can write down the branching ratio to be
\begin{equation}
Br(\mu^+ \rightarrow e^+ \overline{\nu}_e \overline{\nu}_i) =
{|h_{\mu e} f_{ei}|^2|K_{12} K_{32}^*|^2 \over 32 M_1^4 G_F^2}
\left(1-{M_1^2 \over M_2^2}\right)^2\left(1-{M_2^2 \over
M_3^2}\right)^2~.
\end{equation}
Noting that $|K_{12}K_{32}^*| \leq 1/2$, and demanding that $Br$ is in the range
$(0.0015-0.002)$, we obtain an upper limit on the lightest neutral scalar mass
$M_1$: $M_1 \leq (442-412)$ GeV.  If $|K_{12}K_{32}^*|$ is less than $1/2$, $M_1$
will have to be even lighter.  The heavier masses $M_{2}$ and $M_3$ cannot be
very much larger, or else the mixing angle $|K_{12}K_{32}^*|$ will be suppressed.
We estimate $M_{2,3}$ to be not larger than $M_1$ by about a factor of 2.
While these limits hold for the neutral scalars, their charged partners should
also have comparable masses.  Any splitting between the masses of
the neutral and charged scalars will be $SU(2)_L$--breaking, and is limited
by the electroweak $\rho$ parameter.

$\Psi_i$ in Eq. (5) can be $\Psi_e, \Psi_\mu$ or $\Psi_\tau$.  Consider the case when
$\Psi_i = \Psi_e$.  In this case, the Lagrangian of Eq. (5) conserves two combinations
of leptons numbers, viz., $(L_e+3L_\mu)$ and $L_\tau$.  A $Z_3$ subgroup of electron
number is also preserved.  It is not required that these symmetries be
broken, so they may be used to prevent rare processes at undesirable rates.\footnote{We have
in mind a scenario where the solar  and the atmospheric neutrino oscillations arise
from neutrino masses and mixings induced by the seesaw mechanism.  Lepton flavor
violation that arises from the seesaw mechanism is too tiny to be observable in
any processes other than neutrino oscillations themselves.}

If $\Psi_i = \Psi_\mu$ in Eq. (5), then muon and tau lepton numbers are unbroken.
Also unbroken is a $Z_2$ subgroup of electron number.  If $\Psi_i = \Psi_\tau$, then
$(L_e+2L_\mu)$ and $(L_\mu+L_\tau)$ are unbroken, as is $Z_2$ subgroup of $L_e$.
These symmetries guarantee that potentially dangerous lepton number violating
processes remain small.

The scalar fields $\Delta,~\chi$ and $H$ carry lepton number symmetries (or a discrete
subgroup of these symmetries), so that they do not acquire VEVs.  This can be
ascertained by choosing the mass--squared of these scalars to be positive.  The
lepton number symmetries will forbid possible tadpole contributions to their
VEVs.  Thus, these new interactions
do not induce neutrino masses at all.  The only source of neutrino mass in these models
is through the seesaw mechanism.

One may worry  about the compatibility of the lepton number symmetries
present in Eq. (5) and the neutrino oscillation data, which calls for
the breaking of all such
symmetries.  We shall present a concrete example
to demonstrate the consistency.  Consider Eq. (5) with $\Psi_i = \Psi_\mu$.  Let
us impose $L_\mu$ and $L_\tau$ symmetries as well as a $Z_2$ subgroup of $L_e$.
Under these symmetries the scalar fields $(H,~\chi,~
\Delta)$ have charges $(1,0)_-$, $(1,0)_-$ and $(-1,0)_-$ respectively.  The charged
lepton mass matrix
as well as the Dirac neutrino mass matrix will be diagonal due to these symmetries.  Neutrino mixings
can still arise from the superheavy Majorana mass matrix of the $\nu_R$ fields.  The
charges of $(\nu_{eR},~\nu_{\mu R},~\nu_{\tau R})$ under these symmetries are
$\{(0,0)_-; ~(1,0)_+;~(0,1)_+\}$ respectively.  If scalar
fields with charges $(0,0)_+$, $(0,-1)_-$, $(-1,-1)_+$ and $(0,-2)_+$ are introduced
and given large vaccum expectation
values, all neutrino flavors will mix with one another.  These superheavy
scalars will have no couplings to
the $(H,~\chi,~\Delta)$ fields and thus will not spoil the symmetry, except
through small neutrino mass terms.

The exchange of the new scalars with masses of order 500 GeV can lead to new processes.
We list below the most significant of these.

\begin{enumerate}

\item The effective Michel parameter $\rho$ in $\mu$ decay is modified in our scenario:
$(\rho= 0.7485)$.  The deviation from $3/4$ arises because of
the rare
$\mu$ decay mode we have introduced here to explain the LSND data.  Currently
the uncertainty in the measurement of $\rho$ is $\pm 0.0026$, but the TWIST
experiment at TRIUMF \cite{twist} will have a sensitivity of $10^{-4}$ in $\rho$, which
can test this scenario.  There will
also be a small shift  in $G_F$ extracted from $\mu$ decay by $(0.15-0.2)$\%, compared to
the Standard Model value.  Such a shift is currently consistent with radiative
corrections in muon decay (parametrized by $\Delta r$), which has an uncertainty of
about $0.2$\% arising from the $\pm 5$ GeV uncertainty in the top quark mass alone.

\item The neutral component $H^0$ from the scalar $H$ can mediate the process
$e^+ e^- \rightarrow \mu^+ \mu^-$.  The total cross section as well as the
forward--backward asymmetry measured away from the $Z^0$ pole give useful constraints.
The most stringent one arises from LEP experiments run at $130 \le \sqrt{s} < 189$
GeV.  L3 collaboration has quoted lower limits on contact interaction \cite {l3} at
these energies.  The
exchange of $H^0$ induces the following effective contact interaction:
\begin{equation}
{\cal L} = -{|h_{\mu e}|^2 \over 2 m_{H^0}^2} (\overline{\mu}_L \gamma_\mu \mu_L)
\overline{e}_R \gamma^\mu e_R)~.
\end{equation}
This corresponds to the case of $\eta_{LR} = -1$ and all other $\eta_{ij} =0$ in the
notation of Ref. \cite{l3}.  The limit on compositeness scale $\Lambda_- > 1.9$ TeV \cite{l3}
implies
\begin{equation}
m_{H^0}/|h_{\mu e}| \ge 379~GeV~.
\end{equation}
The constraints from forward--backward asymmetry measurements, which have uncertainty
of a few \% will be satisfied once Eq. (9) is met.

The exchange of $\Delta^{++}$ partner of $\Delta^0$ will also contribute to the process
$e^+ e^- \rightarrow \ell^+\ell^-$.  The effective Lagrangian for this process is
\begin{equation}
{\cal L} = {|f_{ei}|^2 \over 2 m_{\Delta^{++}}^2} (\overline{e}_L \gamma_\mu e_L)
(\overline{\ell}_{iL} \gamma^\mu \ell_{iL})~.
\end{equation}
The constraint on $\Lambda_+$ is $\Lambda_+ > \{3.8,~7.3,~3.9\}$ TeV for $\ell_i =
(e,~\mu,~\tau)$ \cite{l3}.  This would lead to the following limits:
\begin{equation}
{m_{\Delta^{++}} \over {|f_{ee}|}} > 758~{\rm GeV},~~
{m_{\Delta^{++}} \over {|f_{e\mu}|}} > 1456~{\rm GeV},~~
{m_{\Delta^{++}} \over {|f_{e\tau}|}} > 778~{\rm GeV}~,
\end{equation}
corresponding to $\ell_i$ being $e,~\mu,$ or $\tau$.
Since $\Delta^{++}$ cannot be split much in mass from $\Delta^0$, explaining the
LSND result at the suggested rate would require $f_{ei}$ to be somewhat smaller
than one and simultaneously one of the neutral scalar masses to be lower than about
300 GeV (see Eqs. (6) and (7)).

\item The charged member $H^+$ from $H$ scalar will mediate the process
$e^+ e^- \rightarrow \nu \nu \gamma$.  This process has been studied at LEP
as a way to count the number of light neutrino species.  L3 collaboration quotes
$N_\nu = 3.05 \pm 0.11 \pm 0.04$ from single photon
measurements carried out at $130 ~{\rm GeV}~ \le \sqrt{s} \le 189$
GeV \cite{l31}.  (Measurements very close to the $Z^0$ pole are less useful for our
purpose.)  Using the 3 sigma limit we obtain (following the procedure outlined
in Ref. \cite{leurer})
\begin{equation}
m_{H^0}/|h_{\mu e}| \ge 375~GeV~.
\end{equation}
We see that the predicted deviation from the Standard model
is in the observable range, once the LSND data is explained.

\item The neutral $H^0$ will contribute to the anomalous magnetic moment
of the muon.  The shift in $a_\mu$ compared to the Standard model prediction
is given by $\delta a_\mu \simeq |h_{\mu e}|^2(m_\mu/m_{H^0})^2/(24 \pi^2)
\simeq 47 \times 10^{-10} |h_{\mu e}|^2(100~{\rm GeV}/m_{H^0})^2$.  This is
in the experimentally interesting range for $|h_{\mu e}|$ being order one
and $m_{H^0} = (100-300)$ GeV, as needed by the LSND data.

\item $\Delta^+$ can mediate $e^+ e^- \rightarrow \nu \nu \gamma$ for the case
where $\Psi_i = \Psi_e$, with a branching ratio very close to the current limits.
In the case where $\Psi_i = \Psi_\mu$, there is new contribution to ordinary
muon decay from the exchange of $\Delta^+$.
We obtain $m_{\Delta^+}/|f_{e\mu}|
\geq 525$ GeV using $\delta(\Delta r) \leq 0.003$.
This limit, for the case where $\Psi_i = \Psi_\mu$, while
consistent, will force some of the other scalars to be lighter than 400 GeV.

\item At a future linear collider running in the $e^- e^-$ mode, it is possible to
produce the doubly charged scalar $\Delta^{++}$ as an s-channel resonance in the
process $e^- + e^- \rightarrow\ell^- \ell^-$, which will provide a spectacular signal.

\item In the dimension 9 operator of Eq. (4), if we insert VEVs to the external
Higgs fields, it becomes a $d=6$ operator.  Then we can attach a $Z^0$ boson line
on the effective four--fermion operator.  That would lead to rare $Z^0$ decays such as
$Z^0 \rightarrow e^+ \mu^- \nu \nu$.  A quick estimate gives the branching ratio
for this decay to be of order $(10^{-7}-10^{-8})$, which may be observable.

\item The $Z^0$ boson can decay into a virtual $H \bar{H}$ pair, which can lead
to $Z^0 \rightarrow e^+ e^- \mu^+ \mu^-$ signal.  This branching ratio is smaller
than $10^{-6}$ .

\end{enumerate}

\subsection{Model for Operator ${\cal O}_2$}

The gauge model that induces operator ${\cal O}_2$ in Eq. (2) can be obtained
in analogous fashion.  The Lagrangian of this model is taken to be
\begin{eqnarray}
{\cal L}_2 &=& h_{e \mu}\overline{\mu}_R \Psi_e H^\dagger + f_{ei} \Psi_e^T C^{-1} \Psi_i
\Delta  \nonumber \\
&+& {\cal M}_0 H^\dagger \chi \Phi + \lambda' \Delta \chi \Phi^\dagger \Phi^\dagger + H.C.
\end{eqnarray}
The main difference from ${\cal L}_1$ is that the helicities of $e$ and $\mu$ have been
switched.  The diagram of Fig. 2 will induce the decay $\mu^+ \rightarrow e^+ +\bar{\nu}_e+
\overline{\nu}_i$.  The phenomenological implications of the model are very similar
to Model 1.  The only significant difference is that the exchange of $H^+$ now cannot
induce the process $e^+ e^- \rightarrow \nu \nu \gamma$, and so the constraint from
that process would not apply.

%%%%%%%%%%%%%%%%%%%%%%%%%%%%%%%%%%%%%%%%%%%%%%%%%%%%%%%%%%%%%%%%%%%
\begin{figure}[h]
\centering
\epsfysize=2.0in
\hspace*{0in}
\epsffile{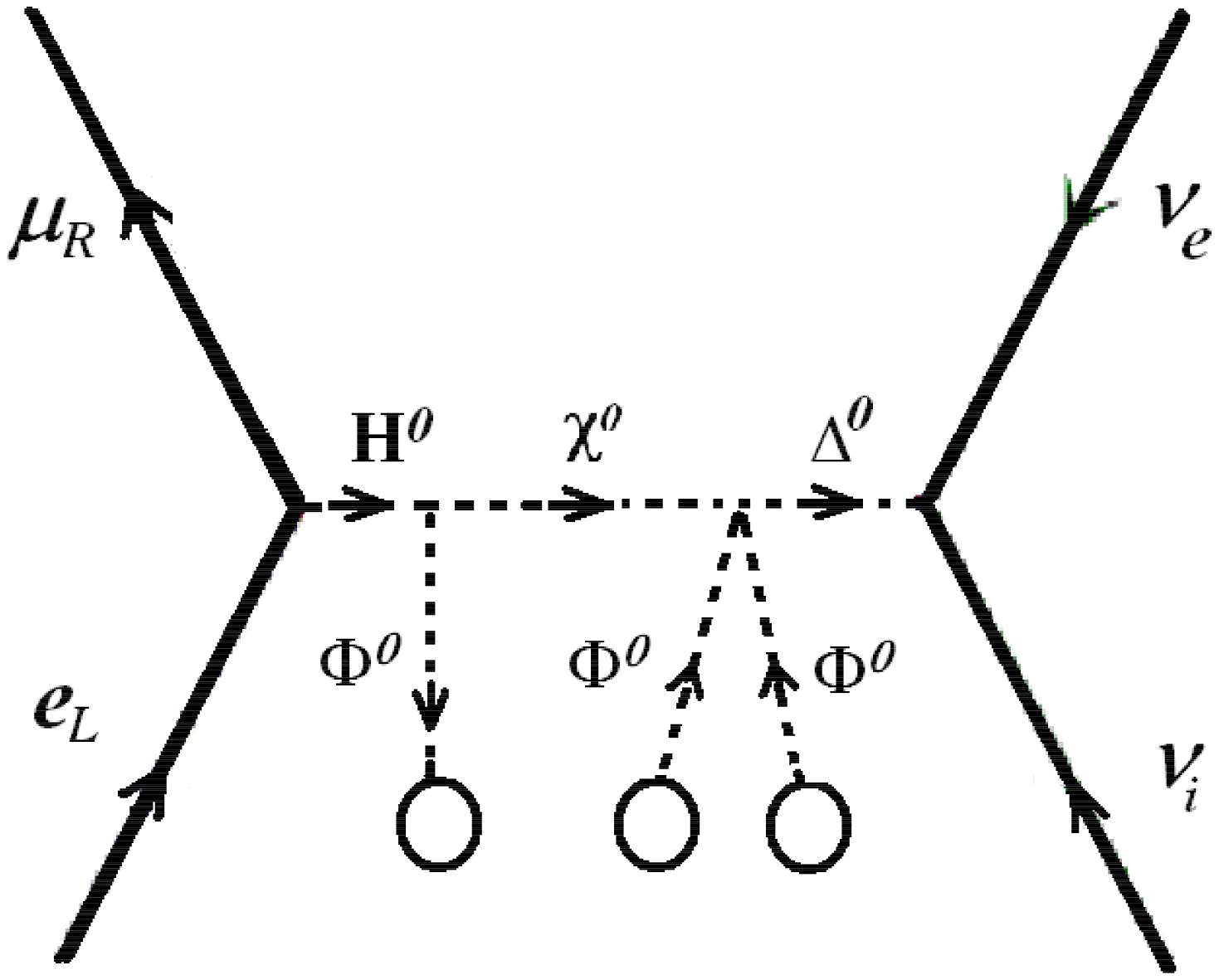}
\caption{Scalar exchange inducing lepton number violating decay
$\mu^+ \rightarrow e^+ + \overline{\nu}_e + \overline{\nu}_i$ through operator ${\cal O}_2$.}
\label{fig2:eps}
\end{figure}
%%%%%%%%%%%%%%%%%%%%%%%%%%%%%%%%%%%%%%%%%%%%%%%%%%%%%%%%%%%%%%%%%%%

\section{Conclusions}

We have presented a class of models where the LSND neutrino anomaly can be explained
in terms of the lepton number violating decay  $\mu^+ \rightarrow e^+ +\bar{\nu}_e+
\overline{\nu}_i$.  There are two effective operators that can lead to this decay.
These operators do not cause any problem with other rare processes such as
$\mu \rightarrow 3 e$, unlike the case of $\Delta L = 0$ decays.  We have
presented gauge models where these operators are derived from renormalizable
interactions.  All models predict the existence of additional scalar fields
with masses below about 500 GeV. Observable deviations in the scattering
processes $e^+ e^- \rightarrow \mu^+ \mu^-,~e^+ e^- \rightarrow \nu \nu \gamma$
and lepton number violating $Z^0$ decays are expected with strengths not much
smaller than the current experimental limits.

Mini--BOONE experiment at Fermilab is expected to go on line in the near future.
Our scenario predicts that mini--BOONE should see no signal.  This is because
the new interactions only affect $\mu^\pm$ decays in our scheme, and not
$\pi^\pm$ decays. Hence, a null result in Mini-Boone does not invalidate the LSND
results.  The proposed
ORLaND experiment \cite{orland} or experiments at a neutrino factory
\cite{bueno} using neutrinos from $\mu$ decays \cite{kuno} can confirm or rule out
the existence of such a lepton number violating decay mode;
with the additional prediction that there be no $L/E$ dependence. The TWIST
experiment at TRIUMF \cite{twist} can test for the shift in the effective Michel
parameter that is predicted in our model.

\section*{Acknowledgments}

We wish to thank K. Eitel for very helpful discussions on
issues related to the analyses of LSND and KARMEN experiments.  It is a
pleasure to thank D. Caldwell,  P. Herczeg, W. Louis,
A. Rubbia and X. Tata for illuminating discussions.
The work of KSB is supported in part by DOE
Grant \# DE-FG03-98ER-41076, a grant from the Research Corporation
and by DOE Grant \# DE-FG02-01ER-45684.
The work of SP is supported in part by DOE Grant \# DE-FG03-94ER40833.
We thank the (Department of Energy's) Institute for
Nuclear Theory at the University of Washington for its hospitality
and the Department of Energy for its partial support during the
completion of this work.

\end{document}